\documentclass[sigconf]{acmart}

\AtBeginDocument{%
  \providecommand\BibTeX{{%
    \normalfont B\kern-0.5em{\scshape i\kern-0.25em b}\kern-0.8em\TeX}}}

\copyrightyear{2024}
\acmYear{2024}
\setcopyright{acmlicensed}\acmConference[ReAQCT '24]{Recent Advances in Quantum Computing and Technology}{June 19--20, 2024}{Budapest, Hungary}
\acmBooktitle{Recent Advances in Quantum Computing and Technology (ReAQCT '24), June 19--20, 2024, Budapest, Hungary}
\acmDOI{10.1145/3665870.3665873}
\acmISBN{979-8-4007-1798-7/24/06}

\usepackage{braket}
\usepackage{amsmath}
\usepackage{physics}

\begin{document}

\title{Suppressing photon detection errors in nondeterministic state preparation}

\author{Csaba Czabán}
\affiliation{%
    \institution{ 
    HUN-REN Wigner Research Centre for Physics}
    \city{Budapest}
    \postcode{H-1525}
    \streetaddress{Konkoly–Thege Miklós út 29-33}
    \country{Hungary}
}
\affiliation{%
    \institution{Eötvös Loránd University, Faculty of Informatics}
    \city{Budapest}
    \country{Hungary}
}
\email{czaban.csaba@wigner.hun-ren.hu}

\author{Zoltán Kolarovszki}
\affiliation{%
\institution{
HUN-REN Wigner Research Centre for Physics}
   \city{Budapest}
    \postcode{H-1525}   \streetaddress{Konkoly–Thege Miklós út 29-33}
    \country{Hungary}
}
\affiliation{%
 \institution{Eötvös Loránd University, Faculty of Informatics}
    \city{Budapest}
    \country{Hungary}
}
\email{kolarovszki.zoltan@wigner.hun-ren.hu}
\author{Márton Karácsony}
\affiliation{%
    \institution{
    HUN-REN Wigner Research Centre for Physics}   \city{Budapest} \postcode{H-1525}
    \streetaddress{Konkoly–Thege Miklós út 29-33}    \country{Hungary}
}
\affiliation{
    \institution{Department of Physics, Duke University}
    \city{Durham, NC} \postcode{NC 27708} \streetaddress{120 Science Dr} \country{USA}
}
\email{karacsony.marton@wigner.hun-ren.hu}

\author{Zoltán Zimborás}
\affiliation{%
   \institution{
   HUN-REN Wigner Research Centre for Physics} \city{Budapest}\postcode{H-1525}\streetaddress{Konkoly–Thege Miklós út 29-33}\country{Hungary}
}
\affiliation{%
   \institution{Eötvös Loránd University, Faculty of Informatics} \city{Budapest}\country{Hungary}
}
\email{zimboras.zoltan@wigner.hun-ren.hu}

\renewcommand{\shortauthors}{Csaba Czabán et al.}

\begin{abstract}
    Photonic quantum computing has recently emerged as a promising candidate for fault-tolerant quantum computing by photonic qubits. These protocols make use of nondeterministic gates, enabling universal quantum computation.
    However, the suggested solutions heavily use particle number resolving detectors (PNRDs), which are experimentally hard to realize and are usually biased in practice. We investigate the possibility of suppressing such errors caused by such photodetector imperfections by adjusting the optimal beamsplitter and phaseshifter angles in the interferometer corresponding to nondeterministic gates. Moreover, we devise an optimization method for determining the adjusted angles, which may achieve higher output state fidelities while controlling the success probabilities of the nondeterministic gates.
\end{abstract}

\ccsdesc[500]{Applied computing~Physics}
\ccsdesc[500]{Information systems~Computing platforms}

\keywords{Quantum optics, quantum computing, simulation, nondeterministic gates, KLM, LOQC}

\maketitle

\section{Introduction}
    Photonic quantum computing has gained traction due to the recent demonstrations of quantum advantage~\cite{photonicadvantage1,zhong2021phase,borealis}. These experiments demonstrated sampling from a distribution, generally accepted as hard using a classical computer. Within these settings, the well-known Gaussian Boson Sampling (GBS) scheme is used, which only uses linear optical quantum gates~\cite{hamilton2017}.

    However, nonlinear couplings are needed between optical modes to perform universal quantum computation. These couplings can be made possible using nonlinear gates (e.g., Kerr gate)~\cite{Azuma_2007,luo2016}. Still, their experimental realization with sufficient strength is difficult, making it difficult to create entangled photonic quantum states in this manner.

    One of the most well-known approaches allowing for universal quantum computation using photonic quantum computers is the Knill-Laflamme-Milburn (KLM) scheme devised in Ref.~\cite{Knill2001}. This protocol creates entanglement between dual-rail-encoded photonic qubits using linear optical gates and photon number resolving detectors (PNRDs). These gates are called nondeterministic gates since the desired photonic qubit operations are realized on successful particle number detection of certain outcomes on the ancilla modes, which makes the setup probabilistic. The scalability of circuits consisting of nondeterministic gates can be achieved by quantum teleportation.

    However, the KLM scheme is tainted by many issues concerning its experimental realization, including the lack of high-quality photon sources or the need for more efficient circuit elements to perform viable error correction~\cite{Pittman_2005}. The latter puts a significant demand on the experimental realization of PNRDs. Correctly detecting photon numbers is critical since the nondeterministic gates may yield an incorrect state after postselecting a faulty measurement outcome.

    Another approach to universal quantum computation with linear optical devices is to use measurement-based quantum computation (MBQC)~\cite{Briegel2009}. The main resources of MBQC are highly-entangled resource states, e.g., cluster states~\cite{Raussendorf2001, Raussendorf2003, NIELSEN2006147}. For this reason, preparing high-fidelity resource states is a critical component of MBQC protocols. Most leading proposals for resource state generation rely on preparing smaller resource states and fusion operations~\cite{Bartolucci2023}. In linear optical quantum computing (LOQC) nondeterministic gates and state preparation via postselection is a promising approach to perform such tasks~\cite{Kieling2007}.

    Meanwhile, photon detection technology with resolved particle numbers has been advancing since the advent of superconducting-nanowire single-photon detectors (SNSPDs)~\cite{SNSPD2001}. An analysis of particle resolution in such a device has been characterized in Ref.~\cite{doi:10.1021/acs.nanolett.3c01228},
    while the effect of imperfect postselection was studied in Ref.~\cite{hamilton2007}. 

    In this article, we explore the possibility of adapting the ideal parameters of nondeterministic gates to suppress the statistical errors arising from the imperfect photon-resolving capabilities of real-life PNRDs.
    Section~\ref{sec:nondeterministic} introduces the general setup of nondeterministic photonic gates with ideal PNRDs, and Sec.~\ref{sec:imperfect} explains the statistical description of postselection with non-ideal PNRDs.
    Finally in Sec.~\ref{sec:suppressing} we detail our quantum state learning-based approach for error suppression in nondeterministic state preparation, similar to methods in Ref.~\cite{Arrazola_2019}.

\section{Nondeterministic photonic gates} \label{sec:nondeterministic}
    
\begin{figure}[ht]
        \centering
        \includegraphics[width = 0.9\linewidth]{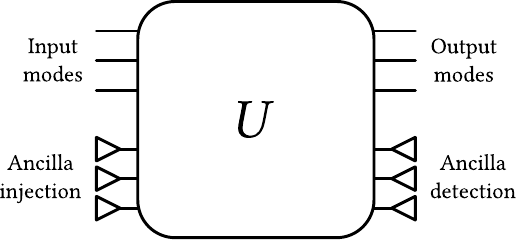}
        \caption{General schematic of a nondeterministic gate. Ancilla photons are injected into additional ancilla modes and channeled into a linear interferometer modeled by a unitary matrix $U$ alongside the input. The output statistics are achieved by postselecting a specific outcome of the ancilla measurement.}
        \Description{}
        \label{fig:nondet-gate}
    \end{figure}

    The typical setup for implementing nondeterministic gates is to prepare states on additional ancilla modes that interfere with the input. Then the ancilla modes are measured using PNRDs (see Fig.~\ref{fig:nondet-gate}). If the measurement outcome matches a specific photon number string, the operation is considered successful and the output can be processed further. Otherwise, the operation fails and the output has to be thrown away.
    
    Consider the setup described in Fig.~\ref{fig:nondet-gate} for a nondeterministic gate and denote the input state as $\rho_{\text{in}}$. Also, let us write the photon number state injected into the ancilla modes as $\ket{\mathbf{a}} = \ket{a_1, a_2, \ldots}$, so that the state entering the linear interferometer is
    \begin{equation}
        \rho = \rho_\text{in} \otimes \ket{\mathbf{a}} \bra{\mathbf{a}}
    \end{equation}
    Then the output of the nondeterministic operation when postselecting on the photon number string $\mathbf{x}$ is
    \begin{equation} \label{eq:output_state}
        \rho_\text{out} = 
        \frac{
            \trace_2 \left\{ U \rho U^\dagger \left(I \otimes \ket{\mathbf{x}} \bra{\mathbf{x}}\right)\right\}
        }{
            S
        }
    \end{equation}
    where $U$ is the unitary action of the linear interferometer on the Fock space. Furthermore, $S$ is the success probability, i.e., the probability of detecting the photon number string $\mathbf{x}$ on the ancilla modes:
    \begin{equation}
        S = \trace \left\{U \rho U^\dagger (I \otimes \ket{\mathbf{x}} \bra{\mathbf{x}})\right\}.
    \end{equation}

    The most well-known example of nondeterministic photonic gates is the conditional sign flip gate introduced in the KLM paper~\cite{Knill2001}.
    While conditional sign flip gate can be optimized analytically \cite{Knill2002}, numerical optimization methods are required for more complicated operations to determine an interferometer configuration with high success probability~\cite{Uskov2010, Gubarev2020, 10.21468/SciPostPhysCore.7.2.032}.
    
\section{Nondeterministic gates with imperfect detectors} \label{sec:imperfect}
    Photon number resolving detectors are critical for implementing nondeterministic gates. In Sec.~\ref{sec:nondeterministic} we assumed perfect particle-resolved detections, however, in practice, particle-resolved detections are imperfect. In this section, we discuss the statistical treatment of the inefficiencies in their photon number resolving capabilities.
    
    A good way to quantify the photon number resolving capabilities of a PNRD is to measure the matrix elements
    \begin{equation} \label{eq:P}
        P_{m,n} \coloneqq \mathbb{P}_\text{readout}(m | n),
    \end{equation}
    which are the conditional probabilities of an $m$ photon readout given that $n$ photons enter the detector. With the $P$ matrix in hand, it is possible to predict the output statistics of nondeterministic gates using the techniques of Ref.~\cite{hamilton2007}.
    
    The off-diagonal entries of the $P$ matrix cause a decrease in the fidelity of the nondeterministic gate. This is because with imperfect detectors the ancilla readout $\mathbf{x}$ no longer guarantees that the output state is given by Eq.~\eqref{eq:output_state}. Hence, the output state will be a mixed state in general. More concretely, we can introduce a new set of conditional probabilities derived from the $P$ matrix
    \begin{equation} \label{eq:cond_prob}
        \mathbb{P}(\mathbf{x} | \mathbf{n}) \coloneqq \prod_{i} \mathbb{P}_\text{readout}(x_i | n_i) = \prod_{i} P_{x_i, n_i},
    \end{equation}
    which are the probabilities of the ancilla readout $\mathbf{x}$ given that the output state is
    $\trace_2 \left\{U \rho U^\dagger \left(I \otimes \ket{\mathbf{n}} \bra{\mathbf{n}}\right)\right\}$ up to normalization.
    For Eq.~\eqref{eq:cond_prob} we assumed that the detectors are independent and identical.
    
    Using the probabilities defined above the output density matrix then can be written as
    \begin{equation}\label{eq:output_state_imperfect}
        \rho_\text{out} = \frac{1}{S} \sum_\mathbf{n} \mathbb{P}(\mathbf{x} | \mathbf{n}) \trace_2 \left\{U \rho U^\dagger \left(I \otimes \ket{\mathbf{n}} \bra{\mathbf{n}}\right)\right\},
    \end{equation}
    where
    \begin{equation} \label{eq:success_prob}
        S = \sum_\mathbf{n} \mathbb{P}(\mathbf{x} | \mathbf{n}) \trace \left\{U \rho U^\dagger \left(I \otimes \ket{\mathbf{n}} \bra{\mathbf{n}}\right)\right\},
    \end{equation}
    is the observed success probability of the nondeterministic gate.

\section{Suppressing PNRD errors by quantum state learning}\label{sec:suppressing}
\begin{figure}[ht]
        \centering
        \includegraphics[width = 0.8\linewidth]{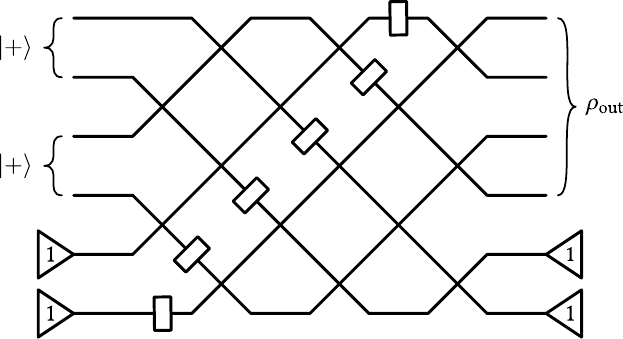}
        \caption{Linear optical circuit used for preparing the resource state $\ket{\psi^*}$ on the output modes. Initially, the two dual-rail qubits are both in the $\ket{+}$ state.}
        \Description{}
        \label{fig:cz-param}
    \end{figure}

    In this section, we investigate the possibility of suppressing the errors in state preparation using the conditional sign flip gate defined in Ref.~\cite{Knill2002} when using such imperfect detectors. More concretely, the goal is to prepare the state
    \begin{equation} \label{eq:resource-state}
        \ket{\psi^*} = \frac{1}{2}\left(
            \ket{0, 1, 0, 1} + \ket{1, 0, 0, 1} + \ket{0, 1, 1, 0} - \ket{1, 0, 1, 0}
        \right)
    \end{equation}
    from Bell states, according to Fig.~\ref{fig:cz-param}, with the highest fidelity possible subject to the restriction that the success probability should be larger then some threshold $S^*$.
    For this, we devise a state learning procedure, similar to the strategy in Ref.~\cite{Arrazola_2019}, but accounting for the bias of the particle-resolved photon detector.

    In more detail, our goal is to train an interferometer to improve the
    output state fidelity when the detectors on the ancilla modes are biased.
    To achieve this desired improvement, we relax the optimal angles in the interferometer, and parametrize the interferometer with beamsplitter and phaseshifter angles from the Clements decomposition~\cite{clements2017optimal}, as demonstrated by Fig.~\ref{fig:cz-param}. We denote the unitary matrix corresponding to the interferometer by $U(\boldsymbol{\theta})$, where $\boldsymbol{\theta}$ represent the angles for the before-mentioned beamsplitters and phaseshifters.
    
    The bias in a particle-resolving SNSPD detector has already been characterized by Resta et. al. in Ref.~\cite{doi:10.1021/acs.nanolett.3c01228}.
    In our experiments, we use this dataset to analyze scenarios involving a maximum of $4$ photons.
    Hence, we only need a part of Table 1. in Ref.~\cite{doi:10.1021/acs.nanolett.3c01228} corresponding to a maximum of $4$ photons, which, according to Eq.~\ref{eq:P}, can be summarized in the
    following detector efficiency matrix:
    \begin{equation} \label{eq:P_Resta}
        P = 
        \begin{pmatrix}
            1.0 & 0.1050 & 0.0110 & 0.0012 & 0.001 \\
            0.0 & 0.8950 & 0.2452 & 0.0513 & 0.0097 \\
            0.0 & 0.0    & 0.7438 & 0.3770 & 0.1304 \\
            0.0 & 0.0    & 0.0    & 0.5706 & 0.4585 \\
            0.0 & 0.0    & 0.0    & 0.0    & 0.4013 \\
        \end{pmatrix}.
    \end{equation}
    Hence, the output state in the before-mentioned setup needs to be calculated according to Eq.~\eqref{eq:output_state_imperfect}, where the detector efficiency matrix $P$ is incorporated. Given $\rho$ initial state density matrix, denote the output state density matrix depending on the interferometer angles $\boldsymbol{\theta}$ as
    \begin{equation}
        \rho_{\mathrm{out}}(\boldsymbol{\theta}) \coloneqq
        \frac{1}{S(\boldsymbol{\theta})} \sum_{\mathbf{n}} \mathbb{P}(\mathbf{x} | \mathbf{n}) \trace_2 \left\{ U(\boldsymbol{\theta}) \rho
        U^\dagger(\boldsymbol{\theta}) \left(I \otimes \ket{\mathbf{n}} \bra{\mathbf{n}}\right)\right\},
    \end{equation}
    where we used the detector efficiency matrix from Eq.~\eqref{eq:P_Resta} to calculate $\mathbb{P}(\mathbf{x} | \mathbf{n})$ according to Eq.~\eqref{eq:cond_prob}, and the success probability $S(\boldsymbol{\theta})$ is given by
    \begin{equation}
        S(\boldsymbol{\theta}) \coloneqq \sum_{\mathbf{n}} \mathbb{P}(\mathbf{x} | \mathbf{n}) \trace \left\{ U(\boldsymbol{\theta}) \rho
          U^\dagger(\boldsymbol{\theta}) \left(I \otimes \ket{\mathbf{n}} \bra{\mathbf{n}}\right)\right\},
    \end{equation}
    according to Eq.~\eqref{eq:success_prob}.

    \begin{figure}[t]
        \centering
        \includegraphics[width = 0.9\linewidth]{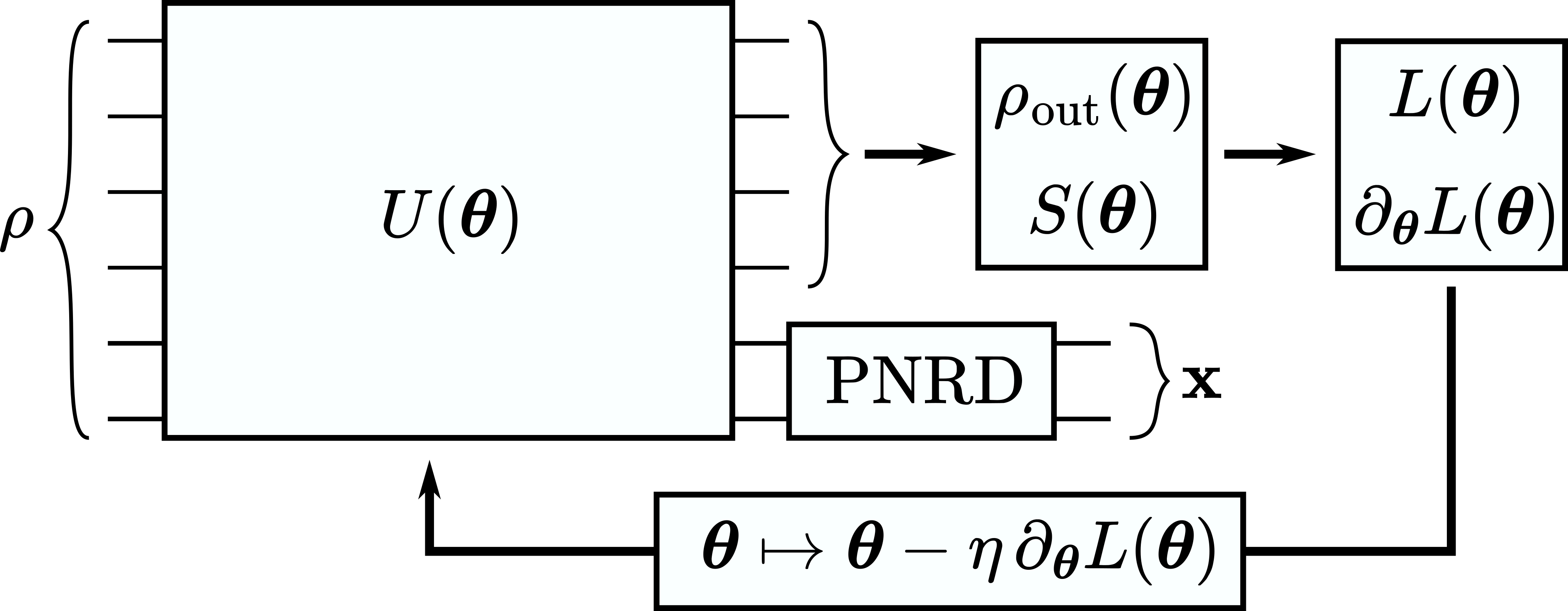}
        \caption{A schematic diagram of the optimization procedure for $4$ modes with $2$ ancilla modes. The Clements decomposition is used to parameterize the possible interferometer configurations. The optimal angles of the beamsplitters and phaseshifters are determined by quantum state learning implemented using Piquasso.}
        \Description{}
        \label{fig:optimization}
    \end{figure}

\begin{figure}[h] \includegraphics[width=0.75\linewidth]{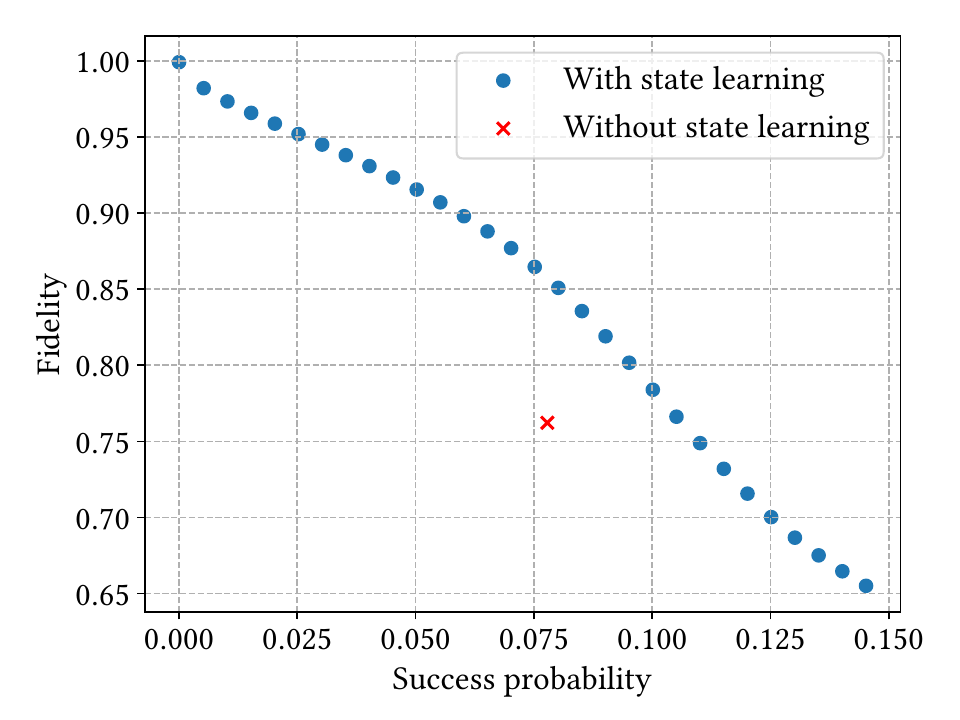}
        \caption{Scatter plot illustrating the fidelity and success probability of the ideal and learned conditional sign flip gate for preparing the state $\ket{\psi^*}$. The diagram shows improvement on the interferometer angles with respect to the original angles optimal for the unbiased scenario.}
        \Description{}
        \label{fig:cz-scatter}
    \end{figure}

  \begin{figure*}[t]
        \centering
        \includegraphics[width=\linewidth]{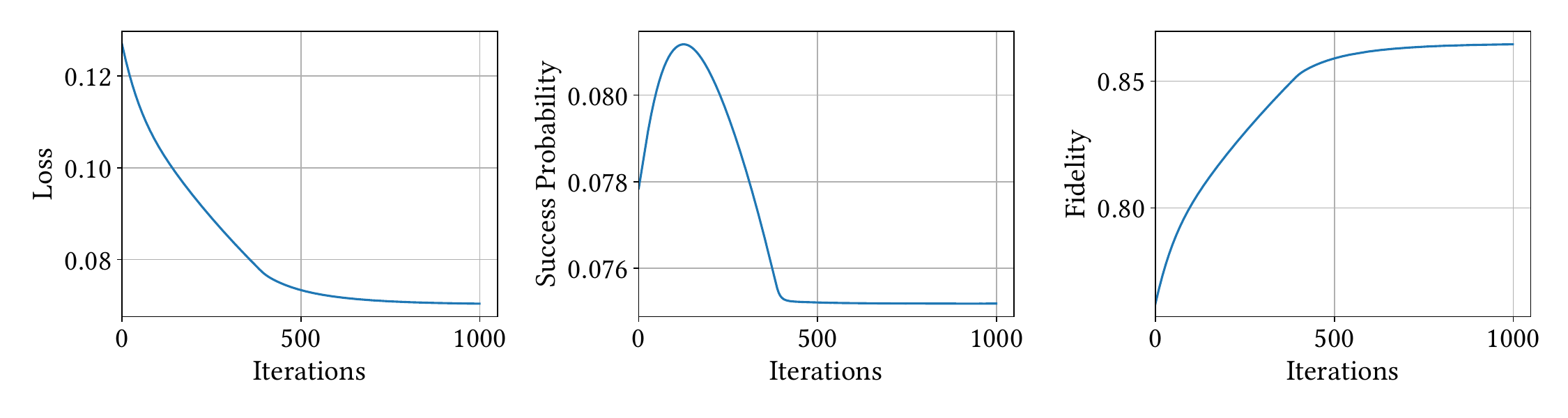}
        \caption{The calculated loss $L(\boldsymbol{\theta})$, 
        the success probability $S(\boldsymbol{\theta})$, and the quantum state fidelity $F(\rho, \sigma) \coloneqq \left(\trace \sqrt{\sqrt{\rho} \sigma \sqrt{\rho}}\right)^2$ produced by the interferometer
        in each iteration during the learning process for the conditional sign flip gate with $S^*=0.075$. The optimization was executed on a 11th Gen Intel(R) Core(TM) i5-1145G7 @ 2.60GHz architecture, where the runtime of $1000$ iterations is approximately $26 s$.}
        \Description{}
        \label{fig:cz-loss-0-075}
    \end{figure*}

    A loss function for state learning tasks has already been defined in Ref.~\cite{Arrazola_2019}. However, for our purposes, this loss function is insufficient, since the success probability from Eq.~\ref{eq:success_prob}
    also needs to be accounted for. Without this, the success probability could arbitrarily decrease in the training procedure.
    To extend the optimization procedure for the success probability, we have employed the $\operatorname{softplus}_\beta$ function defined as
    \begin{equation}
        \operatorname{softplus}_\beta(x) \coloneqq \frac{1}{\beta} \log ( \exp( \beta x ) + 1),
    \end{equation}
    where the parameter $\beta$ controls the ``hardness'', i.e., increasing $\beta$ yields increasingly better approximations of the $\mathrm{ReLU}$ function.
    Finally, the proposed loss function can be written as
    \begin{align}
        L(\boldsymbol{\theta}) &\coloneqq 1 - \sqrt{
            \bra{\psi^*}
            \rho_{\mathrm{out}}(\boldsymbol{\theta})
            \ket{\psi^*}
        }
        + \alpha
        \operatorname{softplus}_\beta (S^* - S(\boldsymbol{\theta})),
    \end{align}
    where $\alpha, \beta, S^*$ are hyperparameters of the training procedure that needs to be chosen depending on the problem. As mentioned previously, $\beta$ controls the hardness of the $\mathrm{softplus}_{\beta}$ function, whereas $\alpha$ can be interpreted as the tradeoff between the contributions coming from the fidelity of the output state and the success rate $S(\boldsymbol{\theta})$. The parameter $S^*$ controls the success probability of the output state. Intuitively, for aptly chosen $\alpha$ and $\beta$, this loss function forces the optimized configuration to produce an output state with a success probability of at least $S^*$.

    We employ the Adam optimizer~\cite{kingma2014adam}, a stochastic gradient descent method, for finding the optimal solution. A schematic diagram of the optimization is shown in Fig.~\ref{fig:optimization}. The interferometer angles are updated according to the gradient of the loss function $L(\boldsymbol{\theta})$, which is calculated using the TensorFlow plugin of Piquasso~\cite{tensorflow2015-whitepaper}. The initial weights $\boldsymbol{\theta}$ are chosen to be the circuit parameters corresponding to the optimal solution with unbiased detectors, and the initial learning rate is $\eta = 0.00025$.
    The state learning tasks were conducted using the Fock simulator of Piquasso, a photonic quantum simulation platform that supports imperfect postselections~\cite{kolarovszki2024piquasso}, and the code is made available at~\cite{script}. The results are summarized in Fig.~\ref{fig:cz-scatter} where the hyperparameters are $\alpha=10$, $\beta=10000$, and $S^*$ ranges from $0$ to $0.1450$. We successfully increased the output fidelity compared to the solution derived with ideal PNRDs. As shown in Fig.~\ref{fig:cz-scatter} it is also possible to achieve near $1$ fidelity at the expense of success probability. The results of a single optimization is illustrated in Fig.~\ref{fig:cz-loss-0-075}.

\section{Conclusion}

    In this work we introduced a state learning algorithm to train linear optical circuits which can prepare resource states nondeterministically. We also demonstrated that the output state fidelity loss due to the photon-number resolving inefficiencies of PNRDs can be suppressed by accounting for their inefficiencies during training.

    Moreover, our state learning method provides explicit control over the success probability of nondeterministic state preparation. This control mechanism allows a reasonable trade-off between the success probability and the output state fidelity; the output state fidelity can be significantly increased without excessively lowering the success probability. As the success probability of a nondeterministic operation is directly proportional to its speed, our algorithm allows fine-tuning optimally the speed and fidelity of nondeterministic state preparation.

    We expect our state learning algorithm to be useful for the experimental realization of universal photonic schemes which rely on the preparation of smaller resource states such as the ones described in Refs.~\cite{Knill2001, kieling_2007, Bartolucci2023}. The results presented in Sec. \ref{sec:suppressing} show that even with current detector technologies it is possible the prepare small high-fidelity resource states when our approach is used to train the optical circuits.

\begin{acks}
This research was supported by the National Research, Development and Innovation Office through the TKP Project no. TKP2021-NVA-04 financed under the TKP2021-NVA funding scheme, the  Quantum Information National Laboratory of Hungary (Grant No. 2022-2.1.1-NL-2022-00004), and grant FK 135220.
ZZ also acknowledges the QuantERA II project HQCC-101017733.
\end{acks}

\bibliographystyle{ACM-Reference-Format}
\bibliography{ref}

\appendix

\end{document}